\documentclass[%
 aip,
 mph,%
 amsmath,amssymb,
 superscriptaddress,
 floatfix,
preprint, 
eqsecnum,%
numerical,%
]{revtex4-1}

\usepackage{graphicx}
\DeclareGraphicsExtensions{.eps,.pdf,.png,.jpg}
\usepackage[caption=false
]{subfig}
\usepackage{epsfig}
\usepackage{epstopdf}
\usepackage{lipsum}
\usepackage{wrapfig}
\usepackage{makecell}
\usepackage{tabularx}
\usepackage{multirow}
\usepackage{soul}
\usepackage{booktabs}
\usepackage{lmodern}
\usepackage{longtable}
\usepackage{amsmath}
\usepackage[OT2,T1]{fontenc}
\usepackage[bookmarks=false]{hyperref}
\usepackage{color}
\usepackage{float}
\hypersetup{pdfauthor={},pdfborder=0 0 0,colorlinks=true,citecolor=blue,linkcolor=blue,urlcolor=blue,}

\DeclareSymbolFont{cyrletters}{OT2}{wncyr}{m}{n}
\DeclareMathSymbol{\comb}{\mathalpha}{cyrletters}{"58}

\newcommand{\ben}{\begin{eqnarray}\displaystyle}
\newcommand{\een}{\end{eqnarray}}

\begin{document}

\title{CBCT scatter correction with dual-layer flat-panel detector}

\author{Xin Zhang}
\thanks{Xin Zhang and Jixiong Xie have made equal contributions to this work and are considered as the first authors.}
\affiliation{Research Center for Advanced Detection Materials and Medical Imaging Devices, Shenzhen Institute of Advanced Technology, Chinese Academy of Sciences, Shenzhen, Guangdong 518055, China} 
\affiliation{University of Chinese Academy of Sciences, Beijing 100049, China}

\author{Jixiong Xie}
\thanks{Xin Zhang and Jixiong Xie have made equal contributions to this work and are considered as the first authors.}
\affiliation{Key Laboratory of Low-grade Energy Utilization Technologies and Systems of Ministry of Education of China, College of Power Engineering, Chongqing University, Chongqing 400044, China}

\author{Yuhang Tan}
\author{Ting Su}
\affiliation{Research Center for Advanced Detection Materials and Medical Imaging Devices, Shenzhen Institute of Advanced Technology, Chinese Academy of Sciences, Shenzhen, Guangdong 518055, China}
\author{Jiongtao Zhu}
\affiliation{Research Center for Advanced Detection Materials and Medical Imaging Devices, Shenzhen Institute of Advanced Technology, Chinese Academy of Sciences, Shenzhen, Guangdong 518055, China} 
\author{Han Cui}
\affiliation{Research Center for Advanced Detection Materials and Medical Imaging Devices, Shenzhen Institute of Advanced Technology, Chinese Academy of Sciences, Shenzhen, Guangdong 518055, China}

\author{Dongmei Xia}
 \thanks{Scientific correspondence should be addressed to Dongmei Xia (xiadm@cqu.edu.cn) and Yongshuai Ge (ys.ge@siat.ac.cn).}
\affiliation{Key Laboratory of Low-grade Energy Utilization Technologies and Systems of Ministry of Education of China, College of Power Engineering, Chongqing University, Chongqing 400044, China}

\author{Hairong Zheng}
\affiliation{Paul C Lauterbur Research Center for Biomedical Imaging, Shenzhen Institute of Advanced Technology, Chinese Academy of Sciences, Shenzhen, Guangdong 518055, China}
\affiliation{Key Laboratory of Biomedical Imaging Science and System, Chinese Academy of Sciences, Shenzhen, Guangdong 518055, China}
\affiliation{National Innovation Center for Advanced Medical Devices, Shenzhen, Guangdong 518131, China.}

\author{Dong Liang}
\affiliation{Paul C Lauterbur Research Center for Biomedical Imaging, Shenzhen Institute of Advanced Technology, Chinese Academy of Sciences, Shenzhen, Guangdong 518055, China}
\affiliation{Key Laboratory of Biomedical Imaging Science and System, Chinese Academy of Sciences, Shenzhen, Guangdong 518055, China}

\author{Yongshuai Ge}
 \thanks{Scientific correspondence should be addressed to Dongmei Xia (xiadm@cqu.edu.cn) and Yongshuai Ge (ys.ge@siat.ac.cn).}
\affiliation{Research Center for Advanced Detection Materials and Medical Imaging Devices, Shenzhen Institute of Advanced Technology, Chinese Academy of Sciences, Shenzhen, Guangdong 518055, China}
\affiliation{Paul C Lauterbur Research Center for Biomedical Imaging, Shenzhen Institute of Advanced Technology, Chinese Academy of Sciences, Shenzhen, Guangdong 518055, China}
\affiliation{Key Laboratory of Biomedical Imaging Science and System, Chinese Academy of Sciences, Shenzhen, Guangdong 518055, China}
\affiliation{National Innovation Center for Advanced Medical Devices, Shenzhen, Guangdong 518131, China.}

\date{\today}

\begin{abstract}
\noindent {\bf Background:} Recently, the popularity of dual-layer flat-panel detector (DL-FPD) based dual-energy cone-beam CT (CBCT) imaging has been increasing. However, the image quality of dual-energy CBCT remains constrained by the Compton scattered X-ray photons.\\
{\bf Purpose:} The objective of this study is to develop an novel scatter correction method, named as e-Grid, for DL-FPD based CBCT imaging.\\
{\bf Methods:} In DL-FPD, a certain portion of the X-ray photons (mainly low-energy primary and scattered photons) passing through the object are captured by the top detector layer, while the remaining X-ray photons (mainly high-energy primary and scattered photons) are collected by the bottom detector layer. Based on the two set of distinct low-energy and high-energy measurements, a linear signal model was approximated for the dual-energy primary and scattered signals on DL-FPD. The distributions of X-ray scatters were quickly estimated using this signal model. Monte Carlo (MC) simulation of a water phantom was conducted to verify this newly developed scatter estimation method. Moreover, physical experiments of water phantom, head phantom, and abdominal phantom were carried out to validate the real performance of this proposed scatter correction method.\\
{\bf Results:} The MC results showed that the e-Grid method was able to generate scatter distributions close to the ground truth. Moreover, the physical experiments demonstrated that the e-Grid method can greatly reduce the shading artifacts in both low-energy and high-energy CBCT images acquired from DL-FPD. On average, the image non-uniformity (NU) was reduced by over 77\% in the low-energy CBCT image and by over 66\% in the high-energy CBCT image. A a consequence, the accuracy of the decomposed multi-material bases was substantially improved. \\
{\bf Conclusion:} The Compton scattered X-ray signals could be quickly corrected using the proposed e-Grid method for DL-FPD based dual-energy CBCT imaging systems.\\
\end{abstract}


\maketitle

\section{Introduction}

Recently, X-ray dual-energy cone-beam computed tomography (DE-CBCT) has attracted considerable research interest. Various data acquisition techniques have been developed to distinguish objects with more than two different materials. For example, the X-ray tube potential modulation (kV switching) technique\cite{muller2016interventional} and the dual-layer flat-panel detector (DL-FPD) based DE-CBCT imaging technique\cite{vlassenbroek2011dual, rassouli2017detector, lu2019dual, ozguner2018objective, shi2020characterization}. Compared to kV switching, DL-FPD enables the acquisition of temporally synchronized dual-energy projections of the object at any gantry rotation angle. Consequently, material-specific CBCT images containing quantitative data are reconstructed. Similar to other FPD based CBCT imaging approach, the DL-FPD based DE-CBCT imaging is also susceptible to Compton scatter due to its large imaging area, e.g., $\ge30$ cm $\times30$ cm. Frequently, Compton scattered X-ray photons induce prominent shading artifacts at the center of reconstructed CBCT images, resulting in a significant degradation of image quality. In DL-FPD based DE-CBCT imaging, these shading artifacts would additionally deteriorate the accuracy of the decomposed basis images. Moreover, the scattered X-ray photons would introduce several other undesired effects to CT images beyond shading artifacts. For instance, streaking artifacts between two dense objects with high attenuation coefficients\cite{joseph1982effects} and pseudo-enhancement of renal cysts\cite{vetter1988correction}.

To date, numerous studies have been conducted aiming to correct the Compton-scattered X-ray signals in CBCT imaging\cite{ruhrnschopf2011general, ruhrnschopf2011general2}. Generally, these methods can be categorized into two groups: hardware-assisted corrections and computation-assisted corrections. Among the hardware-assisted approaches, the placement of an anti-scatter grid\cite{neitzel1992grids} over the entire detector plane stands as the simplest and most widely used method to reject the scattered X-ray photons. Grids with higher ratios can reject more scatter, however, increasing the grid ratio may also result in an increased radiation dose to the patient due to the absorption of primary X-ray photons by the lead strips. The introduction of other devices can also facilitate the CBCT imaging scatter correction. For example, the beam blocker\cite{zhu2009scatter, wang2010scatter, niu2011scatter,lee2012scatter} can be used to measure the amount of scattered X-ray photons on the detector plane. The beam-stop array\cite{love1987scatter,ning2004x, siewerdsen2006simple,zhu2005x} and primary modulator\cite{maltz2006cone, zhu2006scatter, ritschl2015robust} might be utilized to estimate the scatter distribution. Nevertheless, the requirement for two repeated scans may place additional burdens on the current clinical workflow, such as prolonged imaging period\cite{zhu2005x,wang2010scatter}, increased radiation dose, and so on. In addition, increasing the air gap\cite{neitzel1992grids, siewerdsen2000optimization} between the object and the detector can also reduce the scattered X-ray signals. However, this may result in an overall increase of the system's total length if keeping the same imaging field size, i.e., same magnification ratio.

On the other hand, the computation-assisted scatter correction methods include the Monte Carlo (MC) simulation\cite{colijn2004accelerated,
kyriakou2006combining, badal2009accelerating, xu2015practical, sisniega2015high}, kernel-based estimation\cite{hansen1997extraction, ohnesorge1999efficient, maltz2008algorithm, li2008scatter, star2009efficient, sun2010improved}, model-based estimation\cite{meyer2010empirical,zhao2015patient,zhao2016model}, and deep learning approach\cite{maier2018deep,jiang2019scatter, kida2018cone,kurz2019cbct}.
Specifically, MC simulations represent the most accurate approach for scatter estimation and can also be utilized for scatter correction in megavoltage CBCT imaging\cite{spies2001correction}. To enhance computational efficiency, graphics processing units (GPUs) are harnessed to expedite particle transportation calculations\cite{badal2009accelerating}. Furthermore, noise reduction algorithms are employed to optimize simulation time\cite{colijn2004accelerated}. The kernel-based methods, known as scatter kernel superposition (SKS), estimate scatter by convolving the primary signal with a predetermined kernel obtained through MC simulations. The effectiveness of kernel-based SKS approaches has been demonstrated in both kilovoltage diagnostic and megavoltage treatment \cite{maltz2008algorithm} CBCT applications. However, kernel-based methods encounter challenges in accurately capturing variations in object thickness or material composition due to their inherent assumption of symmetric kernel shapes\cite{maltz2008algorithm}. Asymmetric kernels\cite{star2009efficient, sun2010improved} could offer more precise estimations of scatter distribution, though the accuracy heavily depends on the size of the selected kernel segments. The model-based methods estimate the scatter distribution iteratively\cite{meyer2010empirical, zhao2015patient, zhao2016model}, and the performance may also rely on the assumed scatter model and the quality of image segmentation. Most recently, deep learning techniques are employed to obtain the scatter distributions\cite{jiang2019scatter, kida2018cone, kurz2019cbct}. Still, their performance strictly depended on the quality and quantity of training data. As a result, implementing such data-driven approaches in practice is currently challenging, and further investigation is necessary.

In this study, a novel scatter artifact correction method is proposed for DL-FPD based dual-energy CBCT imaging. Unlike the previously mentioned scatter estimation techniques, this approach does not rely on anti-scatter grids or time-consuming computations. The main idea of this new method is to estimate the distributions of scattered X-ray photons through separate measurements of detector responses at two distinct X-ray beam energies. This method is named as e-Grid, which has been demonstrated to be able to significantly reduce the shading scatter artifacts and enhance the accuracy of quantitative dual-energy CBCT imaging.


\begin{figure}[ht]
   \begin{center}
   \includegraphics[width=0.9\linewidth]{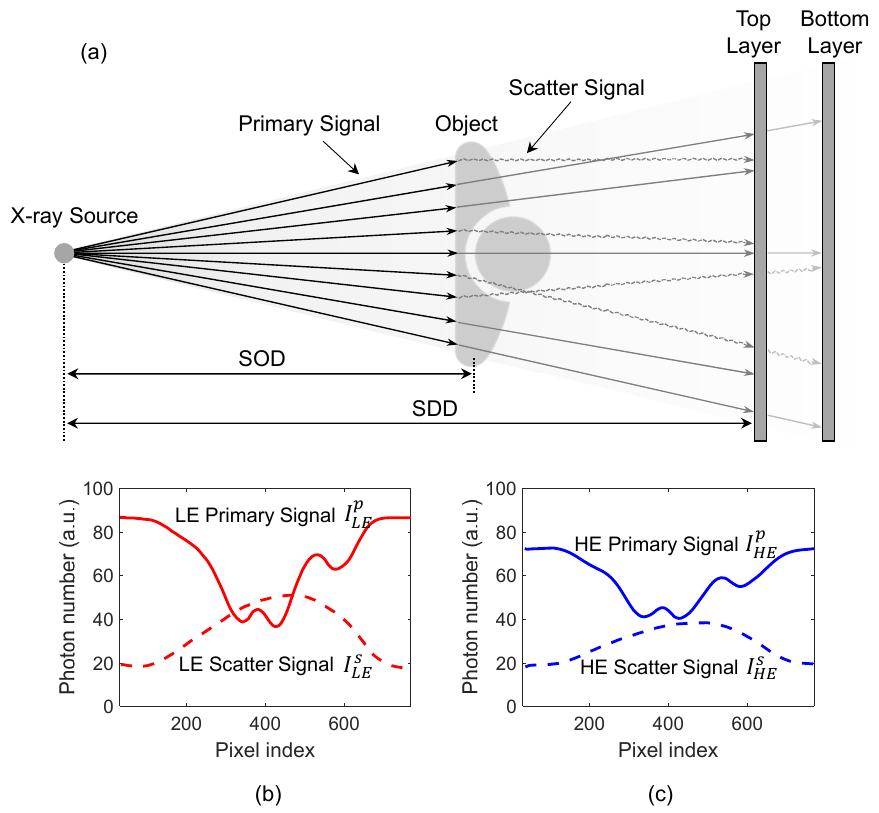}
   \caption{(a) Illustration of a dual-energy CBCT imaging system with a DL-FPD. (b) The detected low-energy primary and scattered X-ray signals by the top detector layer. (c) The detected high-energy primary and scattered X-ray signals by the bottom detector layer. Optionally, a Copper filter can be placed between the top and bottom detector layers to increase the energy separation.}
   \label{fig:schematic}
    \end{center}
\end{figure}

\section{Methods and Materials}\label{sec: method}
\subsection{Signal model}
In the proposed e-Grid approach, it was assumed that the X-ray intensities detected by each layer of the DL-FPD consist of two components: the primary X-ray signal and the Compton scattered X-ray signal, see Fig.~\ref{fig:schematic}. In general, the primary X-ray signal refers to the X-ray beam that travels directly from the X-ray source to the detector without being scattered or absorbed by the object, while the Compton scattered X-ray photons were those that had been deflected away from their original path due to interactions with the object. Mathematically, the low-energy (LE) signal intensity $I_{LE}$ and the high-energy (HE) signal intensity $I_{HE}$ were expressed in terms of the following linear formulas:

\begin{align}
I_{LE}&=I^{p}_{LE}+I^{s}_{LE},\label{eq:1}\\
I_{HE}&=I^{p}_{HE}+I^{s}_{HE},\label{eq:2}
\end{align}
where $I^{p}_{LE}$ and $I^{p}_{HE}$ denoted the primary low-energy signal intensity and the primary high-energy signal intensity, respectively, $I^{s}_{LE}$ and $I^{s}_{HE}$ denoted the scattered low-energy signal and the scattered high-energy signal, respectively. To retrieve the scattered signals, the following approximations were assumed,

\begin{align}
I^{p}_{LE}&=f_p(I^{p}_{HE})\approx\alpha^p_{1}I^{p}_{HE}+\alpha^p_{0},\label{eq:3}\\
I^{s}_{LE}&=f_s(I^{s}_{HE})\approx \alpha^s_{1}I^{s}_{HE}+\alpha^s_{0},\label{eq:4}
\end{align}
where function $f_p$ and $f_s$ were assumed to map the high-energy signals onto the low-energy signals\cite{brunner2011prior}. Moreover, functions $f_p$ and $f_s$ were approximated by linear expansions with first-order coefficients $\alpha^p_{1}$, $\alpha^s_{1}$ and zero-order coefficients $\alpha^p_{0}$, $\alpha^s_{0}$. By
 substituting Eq.~(\ref{eq:3}) and Eq.~(\ref{eq:4}) into Eq.~(\ref{eq:1}), one gets:

\begin{align}
I_{LE}&\approx \alpha^p_{1}I^{p}_{HE}+\alpha^s_{1}I^{s}_{HE}+\alpha^p_{0}+\alpha^s_{0}, \label{eq:5}
\end{align}
By jointly solving Eq.~(\ref{eq:2}) and Eq.~(\ref{eq:5}), the high-energy scatter signal $I^{s}_{HE}$ was found equal to:
\begin{align}
I^{s}_{HE}=\frac{I_{LE}-\alpha^{p}_{1}I_{HE}-\alpha^{p}_{0}-\alpha^{s}_{0}}{\alpha^s_{1}-\alpha^p_{1}}.\label{eq:7}
\end{align}
Substituting Eq.~(\ref{eq:7}) into Eq.~(\ref{eq:4}), eventually, the low-energy scatter signal $I^{s}_{LE}$ can also be determined.

\begin{figure}[t]
   \begin{center}
   \includegraphics[width=0.95\linewidth]{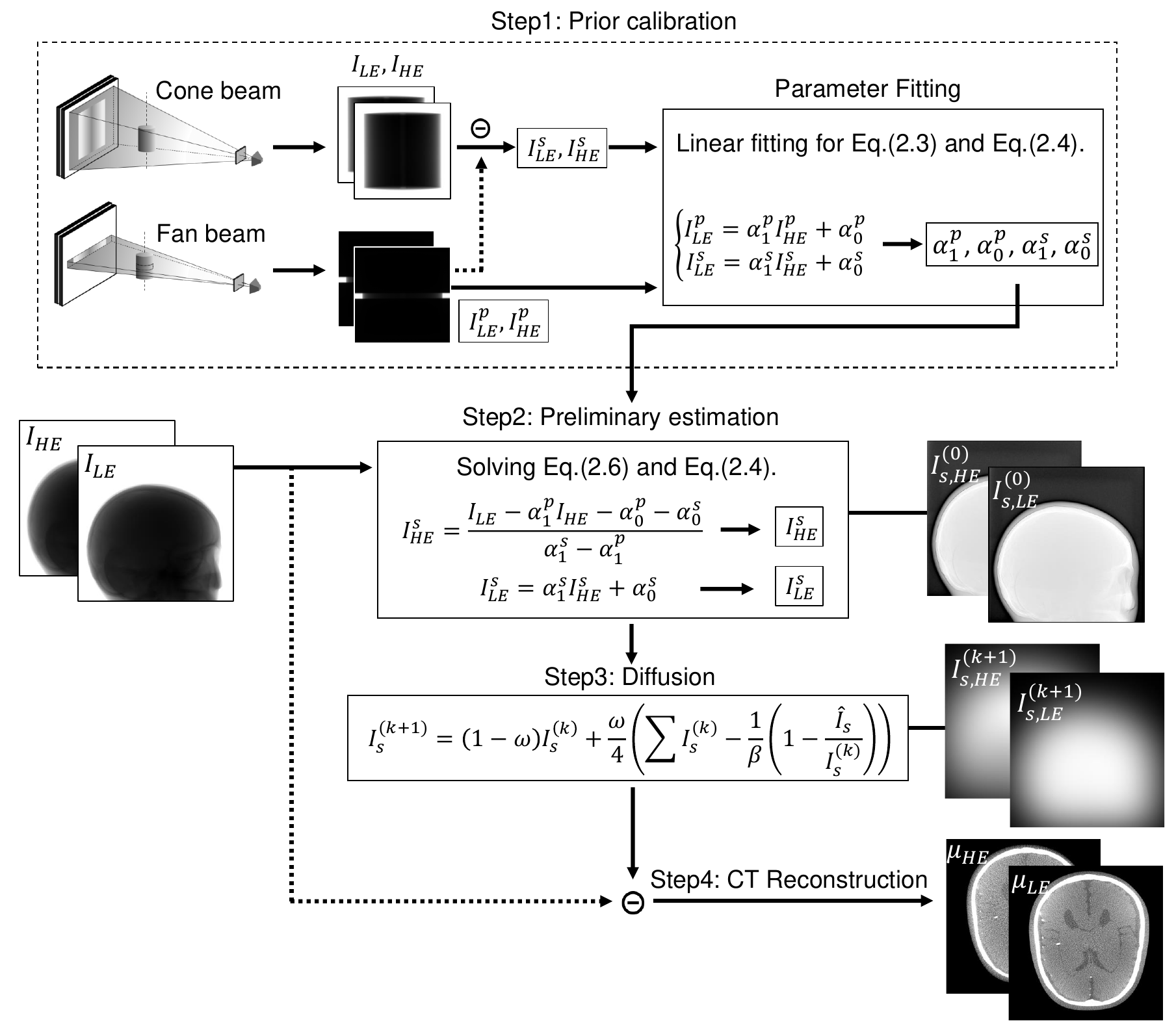}
   \vspace*{-2ex}
   \caption{The workflow of the proposed scatter correction method with DL-FPD. In the first step, parameters $\alpha^p_{1}$, $\alpha^p_{0}$, $\alpha^s_{1}$ and $\alpha^s_{0}$ were calibrated based on phantom measurements obtained from cone beam and fan beam imaging. In the second step, distributions of the scattered signals from the two detector layers were calculated, followed by diffusion of the high-frequency components in the third step. Finally, CBCT images without shading artifacts were reconstructed in the fourth step.}
   \label{fig:workflow}
    \end{center}
\end{figure}

\subsection{Signal estimation}
The parameters $\alpha^p_{1}$, $\alpha^p_{0}$, $\alpha^s_{1}$ and $\alpha^s_{0}$ in Eq.~(\ref{eq:3}) and Eq.~(\ref{eq:4}) need to be determined in prior before estimating the scatter distributions. To do so, certain calibration phantom had to be scanned with a cone beam (23 cm beam width on the detector plane) and a fan beam (0.5 cm beam width on the detector plane), separately. The proposed calibrations were illustrated in Step~1 in Fig.~\ref{fig:workflow}. Both primary and scattered signals were measured in the cone beam imaging experiment, whereas, only the primary signal ($I^{p}_{LE}$ and $I^{p}_{HE}$) was measured in the fan beam imaging experiment. Herein, it was assumed that the scattered signals were negligible in the fan beam imaging experiment. Afterwards, parameters $\alpha^p_{1}$, $\alpha^p_{0}$, $\alpha^s_{1}$ and $\alpha^s_{0}$ were determined through linear fitting. 

In this work, a polymethyl methacrylate (PMMA, mass density of $\rm{1.19~g/cm^3}$) cylinder phantom with a diameter of 16 cm was calibrated for head imaging, while a water cylinder phantom with a diameter of 30 cm was calibrated for body imaging. It should be noted that the accuracy of the calibrations and linear fittings had a crucial impact on the final performance of the e-Grid method. In practice, some high-frequency residual structures remain in the retrieved $I^{s}_{LE}$ and $I^{s}_{HE}$ signals. To remove these high-frequency components, an iterative diffusion algorithm\cite{jia2012gpu} was employed, namely:

\begin{align}
I_{s}^{(k+1)}(u,v)&=(1-\omega)I_{s}^{(k)}(u,v)+\frac{\omega}{4}\Bigl(\sum{I_{s}^{(k)}(u,v)}-\frac{1}{\beta}\left(1-\frac{\hat I_{s}(u,v)}{I_{s}^{(k)}(u,v)}\right)\Bigl),\label{eq:Diff}
\end{align}
where (${u}$, ${v}$) denoted the pixel index, $\sum{I_{s}^{(k)}(u,v)} = I_{s}^{(k)}(u-1,v) + I_{s}^{(k)}(u,v-1) +I_{s}^{(k)}(u+1,v) + I_{s}^{(k)}(u,v+1)$, $\hat{I}_{s}(u,v)$ denoted the estimated low-energy signal or high-energy scatter signal, $k$ denoted the iteration step. In our studies, $\omega = 0.8$ and $\beta=1 \times 10^3$ were utilized. The entire workflow of the proposed scatter correction procedure was illustrated in Step~2 in Fig.~\ref{fig:workflow}. 

\subsection{Monte Carlo simulation}
Monte Carlo (MC) simulations were conducted on GATE (ver 9.2) platform\cite{jan2004gate, sarrut2021advanced} to validate the feasibility of this newly proposed scatter correction method. A 10 cm diameter digital water phantom containing two 2.5 cm diameter inserts filled with iodine solution of 20 mg/ml concentration was imaged. To mimic the physical experiments, a DL-FPD was simulated: the CsI scintillator on the top detector layer was 0.26 mm thick, and the  CsI scintillator on the bottom layer was 0.55 mm thick. An additional 1.0 mm copper filter was inserted between the two detector layers. Each detector layer had $190 \times 100$ detector elements with 0.616~mm $\times$ 0.616~mm element dimension. The incident beam spectrum, which was the same as used for physical experiment, used in GATE was generated at 125 kV tube potential with 0.4 mm Cu filtration. The source to detector distance (SDD) was fixed at 1200 mm, the source to iso-center distance (SOD) was fixed at 1130 mm. More details of the MC simulation were listed in Table~\ref{tab:TG-268} and Table~\ref{tab:parameter}.

\begin{table}[!htb]
\begin{center}
\caption{Key setups of the Monte Carlo simulation following the AAPM TG-268 protocol\cite{sechopoulos2018records}.
\vspace*{0ex}
}
\renewcommand\arraystretch{1}
\resizebox{1\columnwidth}{!}{
\begin{tabular} {p{6cm}p{10cm}}
\toprule[1pt]
\hline
Item&Description\\
\cline{1-2}
Software&GATE v9.2 (Geant4\cite{agostinelli2003geant4} v11.1.1 and Root\cite{brun1997root} v6.26). \\
Hardware&Intel Xeon(R) Gold 6248R CPU @ 3.00GHz.   \\
Physics and transport &The simulated physics was managed by the Geant4 Monte Carlo kernel, which was responsible for tracking particles in matter and processing physical interactions.    \\
Histories statistical uncertainty & $2.35\times 10^{11}$ events per projection.    \\
Timing& Approximately 3750 seconds to run 1 projection.  \\
Scored quantities &X-ray photon deposition events. \\
\hline
\bottomrule[1pt]
\end{tabular}}
\label{tab:TG-268}
\end{center}
\end{table}

\begin{table}[h]
\begin{center}
\caption{The key imaging parameters used for MC simulations and experiments.
\vspace*{2ex}
}
\renewcommand\arraystretch{1}
\resizebox{0.5\columnwidth}{!}{
\begin{tabular} {lcc}
\toprule[1pt]
\hline
  &MC& Experiments\\
\cline{1-3}
Detector array   &190$\times$100& 768$\times$768\\
Detector element size (mm) &0.616 & 0.308 \\
Projection views &360& 450\\
Tube potential (kV) &125 & 125\\
Tube current(mA) &-& 7.1 \\
Beam filtration: Cu (mm) &0.4 & 0.4\\
Extra filtration: Cu (mm) &1.0& 1.0\\

\hline
\bottomrule[1pt]
\end{tabular}}
\label{tab:parameter}
\end{center}
\end{table}

\begin{figure}[htbp]
	\begin{center}
		\includegraphics[width=0.8\linewidth]{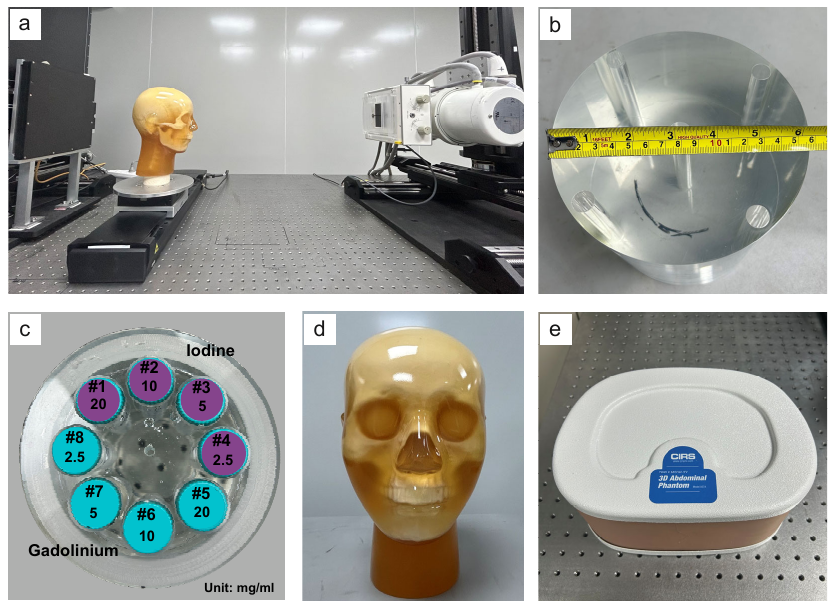}
		\caption{(a) The experiment benchtop system, (b) the PMMA cylinder phantom, (c) the water phantom with inserts of iodine and gadolinium solutions, (d) the head phantom, and (e) the abdominal phantom.}
		\label{fig:sample}
	\end{center}
\end{figure}

\subsection{Physical experiment}
Phantom experiments were conducted on our benchtop system, which was equipped with a medical-grade X-ray tube (G-242, Varex, USA) and a dual-layer FPD (560RF-DE, Careray, China), see Fig.~\ref{fig:sample}(a). The X-ray tube voltage was fixed at 125 kV with 12.5 mA tube current, and the X-ray beam was filtered by 0.4 mm Copper. The DL-FPD had $1536 \times 1536$ detector elements with a native element dimension of 0.154~mm $\times$ 0.154~mm, corresponding to a field-of-view (FOV) of 23.65 cm$\times$23.65 cm. During data acquisition, the $2 \times 2$ binning mode was applied for DF-FPD. The CsI:Tl material was 0.26 mm thick on the top detector layer and 0.55 mm thick on the bottom detector layer. The top detector layer and bottom detector layer were 6.6~mm apart, and an additional 1.0 mm copper filtration was inserted between them to increase spectrum separation. 

Three phantoms were scanned: a self-made water cylinder of 10 cm diameter, see Fig.~\ref{fig:sample}(c); an angiographic head phantom (Model: 41309-300, Kyoto Kagaku, Japan) of 16 cm diameter, see Fig.~\ref{fig:sample}(d); an abdominal phantom (Model: 057A, CIRS, USA) of 28 cm diameter, see Fig.~\ref{fig:sample}(e). Inside the water cylinder, tubes 1 to 4 contain iodine solution with concentration of 20 mg/ml, 10 mg/ml, 5 mg/ml, and 2.5 mg/ml, respectively. Moreover, tubes 5 to 8 were filled with gadolinium solution with concentrations of 20 mg/ml, 10 mg/ml, 5 mg/ml, and 2.5 mg/ml, respectively. The source to detector distance (SDD) was fixed at 1200 mm, and the source to iso-center distance (SOD) was set at 950 mm, 1050 mm and 910 mm for the water phantom, head phantom, and abdominal phantom, respectively. Adjusting the SOD enabled us to position the object as close to the detector as possible, resulting in the collection of a sufficient number of Compton scatters. More details of the experiment setups were provided in Table~\ref{tab:parameter}. Note that the experimental setup and MC setup are not identical. For example, the MC utilized lower spatial resolution and a smaller FOV than the actual experiment. The current settings for MC were chosen solely to save the computation time in GATE. To image the abdominal phantom, the DL-FPD was laterally shifted at two positions, each covering more than half of the phantom. Afterwards, the projections were stitched correspondingly to generate the dual-energy imaging data of the abdominal phantom.

\subsection{Evaluation metric}
To quantify the correction performance, the image non-uniformity (NU) indices were measured. Explicitly, NU was defined as follows: 
\begin{equation}
\text{NU}=\left|\frac{{\bar{\mu}_c}-{\bar{\mu}_e}}{{\bar{\mu}_e}}\right|\times 100\%,\label{eq:nu}
\end{equation}
where ${\bar{\mu}_c}$ and ${\bar{\mu}_e}$ denoted the mean value of the region of interest (ROI) selected at the center and at the edge of the reconstructed CBCT images, respectively.

In addition, the signal-to-noise ratio (SNR) of the CT images was compared before and after scatter correction. The SNR of the CT images was defined as:
\begin{equation}
\text{SNR}=\frac{{\bar{\mu}_{\rm{s}}}}{{\sigma_{\rm{n}}}},\label{eq:snr}
\end{equation}
where $\bar{\mu}_{\rm{s}}$ denoted the mean value of the selected ROI on the reconstructed CT images, and $\sigma_{\rm{n}}$ denoted the corresponding standard deviation.

\section{Results}\label{sec: results}
\subsection{MC results}
\begin{figure}[htbp]
	\begin{center}
		\includegraphics[width=0.95\linewidth]{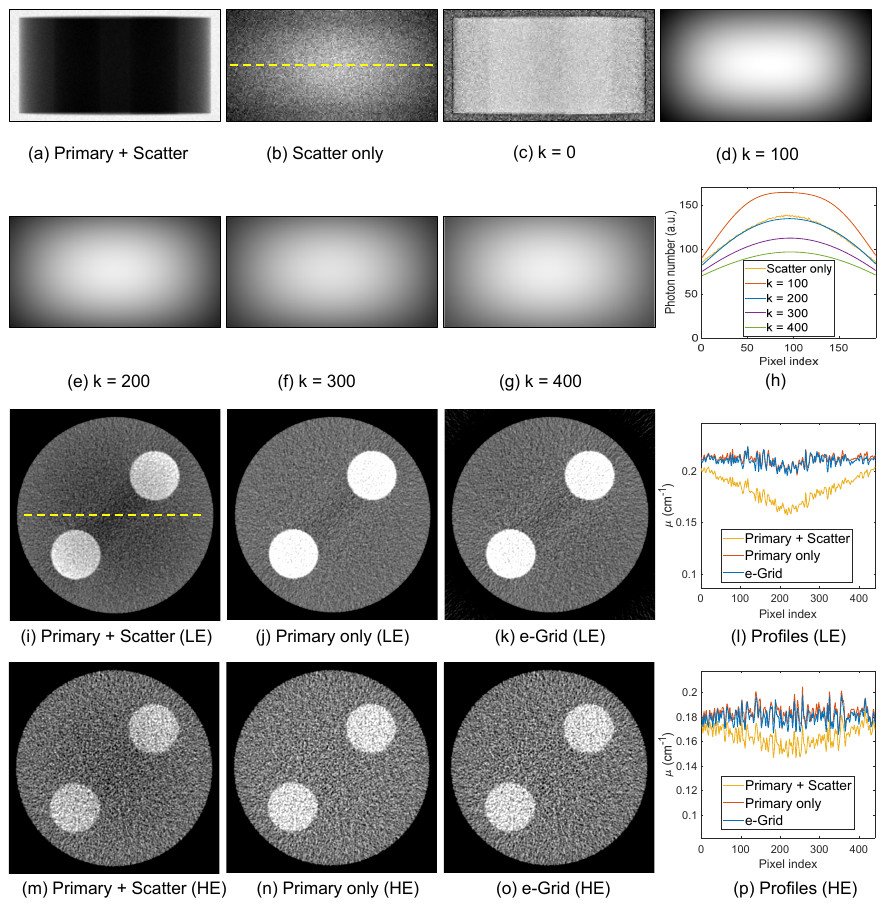}
		\caption{The MC simulation results, (a) total signal, (b) scatter signal, (c) the calculated scatter signal before diffusion, (d)-(g) diffused projections with k=100, 200, 300 and 400 iterations, respectively, (h) line profile comparison results, (i)-(l) the low-energy CBCT imaging results, (m)-(p) the high-energy CBCT imaging results. The display window was $\left[0.11, 0.36\right]$ $\rm{cm}^{-1}$ for the low-energy CT images, and $\left[0.09, 0.29\right]$ $\rm{cm}^{-1}$ for the high-energy CT images.} 
		\label{fig:MC_scatter}
	\end{center}
\end{figure}
The MC simulation results were presented in Fig.~\ref{fig:MC_scatter}. The projection containing total (primary and scatter) signals at a certain angle was presented in Fig.~\ref{fig:MC_scatter}(a), while the scatter only signal $I^s_{LE}$ was presented in Fig.~\ref{fig:MC_scatter}(b). As seen, the scatter signal mainly consists of low-frequency components. However, the scatter signal $I^{s,k=0}_{LE}$ estimated from Eq.~(\ref{eq:4}) and Eq.~(\ref{eq:7}) contained clear residual structures of high frequency, see Fig.~\ref{fig:MC_scatter}(c). The scatter distributions processed with Eq.~(\ref{eq:Diff}) at four different iteration steps ($k=$ 100, 200, 300, and 400) were presented in Fig.~\ref{fig:MC_scatter}(d)-(g), respectively. The line profiles in Fig.~\ref{fig:MC_scatter}(h) demonstrated that diffusion can smooth out the structural residuals and generated a distribution similar to the ground truth at approximately 200 iterations.

The low-energy CT images reconstructed from the total signal, primary signal (ground truth), and scatter corrected signal were shown in Fig.~\ref{fig:MC_scatter}(i)-(k), respectively. The high-energy CT images reconstructed from the total signal, primary signal (ground truth), and scatter corrected signal were shown in Fig.~\ref{fig:MC_scatter}(m)-(o), respectively. Visually, the CBCT image reconstructed before scatter correction exhibits noticeable shading artifacts in the central region, indicating the presence of strong scatter artifacts. These shading artifacts were significantly reduced after processing with the proposed e-Grid method. Comparing to the ground truth, the e-Grid method produced similar results, as depicted in the profiles shown in Fig.~\ref{fig:MC_scatter}(l) and Fig.~\ref{fig:MC_scatter}(p). On the low-energy CT images, minor beam hardening artifacts were observed between the two iodine inserts.
\subsection{Experimental results}
\begin{figure}[htbp]
\begin{center}
\includegraphics[width=0.95\linewidth]{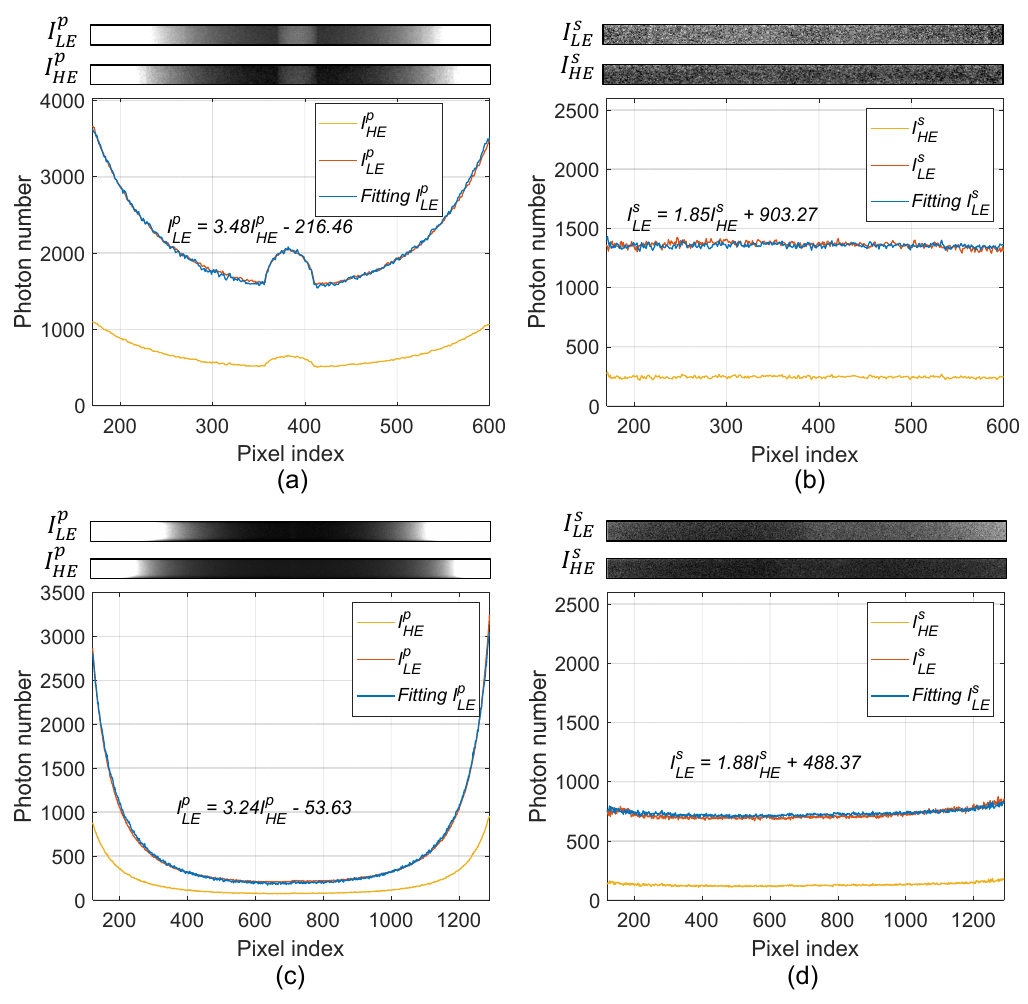}
\caption{Calibration results of the 16 cm PMMA phantom (a) the primary signal, $\alpha^p_{1}=3.48$, $\alpha^p_{0}=-216.46$, (b) the scattered signal, $\alpha^s_{1}=1.85$, $\alpha^s_{0}=903.27$. Calibration results of the 30 cm water phantom (c) the primary signal, $\alpha^p_{1}=3.24$, $\alpha^p_{0}=-53.63$, (d) the scattered signal, $\alpha^s_{1}=1.88$, $\alpha^s_{0}=488.37$.}
\label{fig:alpha_fitting}
\end{center}
\end{figure}
\begin{figure}[t]
   \begin{center}
   \includegraphics[width=1\linewidth]{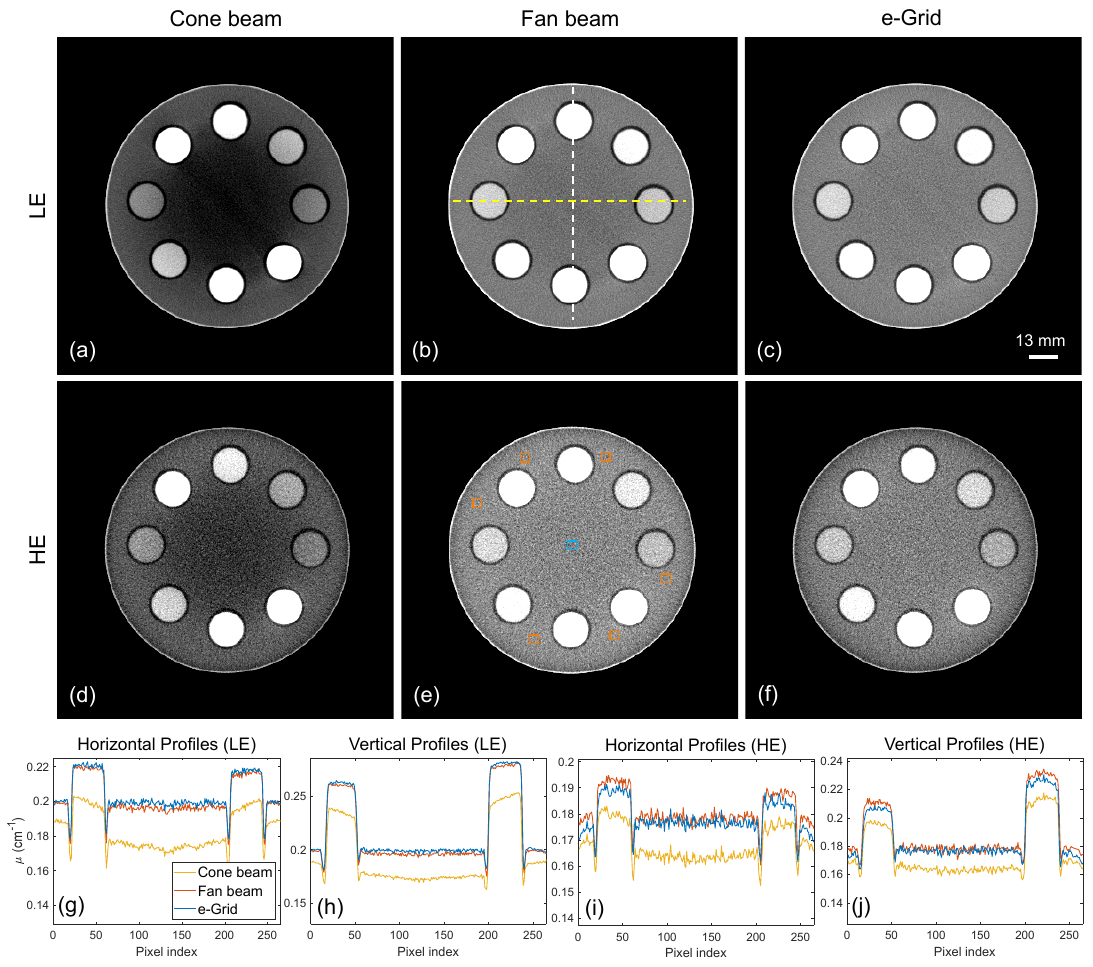}
   \caption{Imaging results of the water cylinder phantom. The profiles along the horizontal and vertical directions were compared and plotted at the bottom. The display window was $\left[0.17, 0.23\right]$ $\rm{cm}^{-1}$ for low-energy CT, and $\left[0.15, 0.20\right]$ $\rm{cm}^{-1}$ for high-energy CT. The scale bar denoted 13 mm.} 
   \label{fig:sc_results_insert}
    \end{center}
\end{figure}

The experimental calibration results were shown in Fig.~\ref{fig:alpha_fitting}. In particular, the calibration coefficients from the 16 cm PMMA phantom were used by e-Grid to correct the dual-energy CBCT imaging results of the water and head phantoms, while the calibration coefficients from the 30 cm water phantom were used by e-Grid to correct the dual-energy CBCT imaging results of the abdominal phantom.

The experimental results of the water phantom were shown in Fig.~\ref{fig:sc_results_insert}. The low-energy CT images were presented in the first row, while the high-energy CT images were presented in the second row. From left to right, the CT images were generated from cone beam, fan beam, and e-Grid (cone beam with scatter correction), respectively. As seen, CT images obtained from cone beam exhibit pronounced shading artifacts before scatter correction, particularly noticeable in the low-energy CT image. Results found that narrowing the width of the X-ray beam can greatly mitigate the scatter shading artifacts. Quantitatively, the measured scatter-to-primary ratio (SPR)\cite{bootsma2011effects}, defined as the ratio of scattered X-rays to primary X-rays, was less than 7\% for the two detector layers in the fan beam experiment. Visually, the proposed e-Grid method can eliminate the scatter artifacts in both low-energy and high-energy CBCT images, resulting in a more uniform signal distribution, see the horizontal and vertical line profile results in Fig.~\ref{fig:sc_results_insert}(g)-(j).

\begin{figure}[htbp]
	\begin{center}
		\includegraphics[width=1\linewidth]{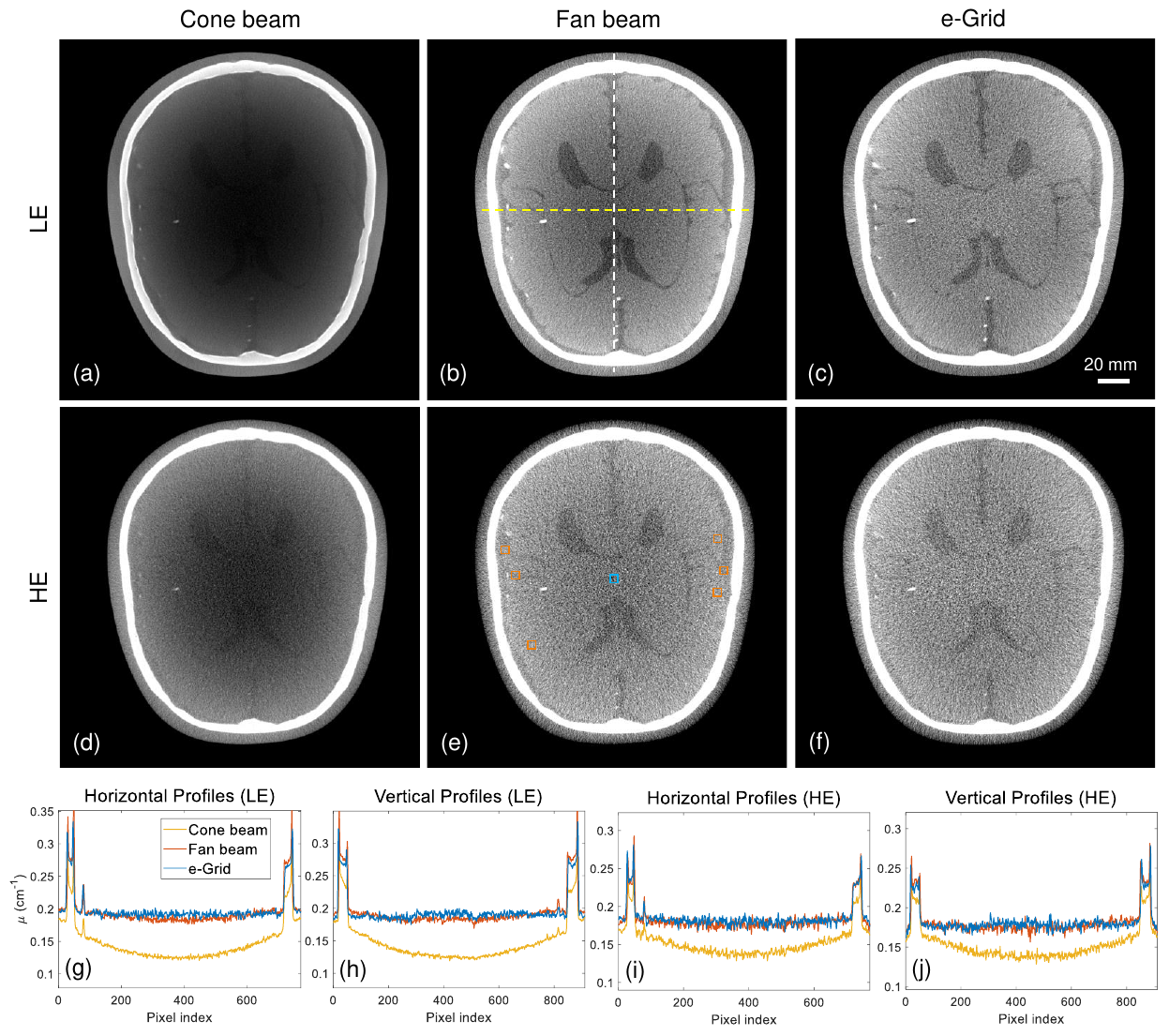}
		\caption{Imaging results of the head phantom. The profiles along the horizontal and vertical directions were compared and plotted at the bottom. The display window was $\left[0.12, 0.25\right]$ $\rm{cm}^{-1}$ for low-energy cone beam CT images, and $\left[0.17, 0.21\right]$ $\rm{cm}^{-1}$ for low-energy fan beam and e-Grid CT images. The display window was $\left[0.13, 0.20\right]$ $\rm{cm}^{-1}$ for high-energy cone beam CT images, and $\left[0.16, 0.20\right]$ $\rm{cm}^{-1}$ for high-energy fan beam CT images and e-Grid processed CT images. The scale bar denoted 20 mm.} 
		\label{fig:sc_results_KYOTO}
	\end{center}
\end{figure}

\begin{figure}[htbp]
   \begin{center}
   \includegraphics[width=0.9\linewidth]{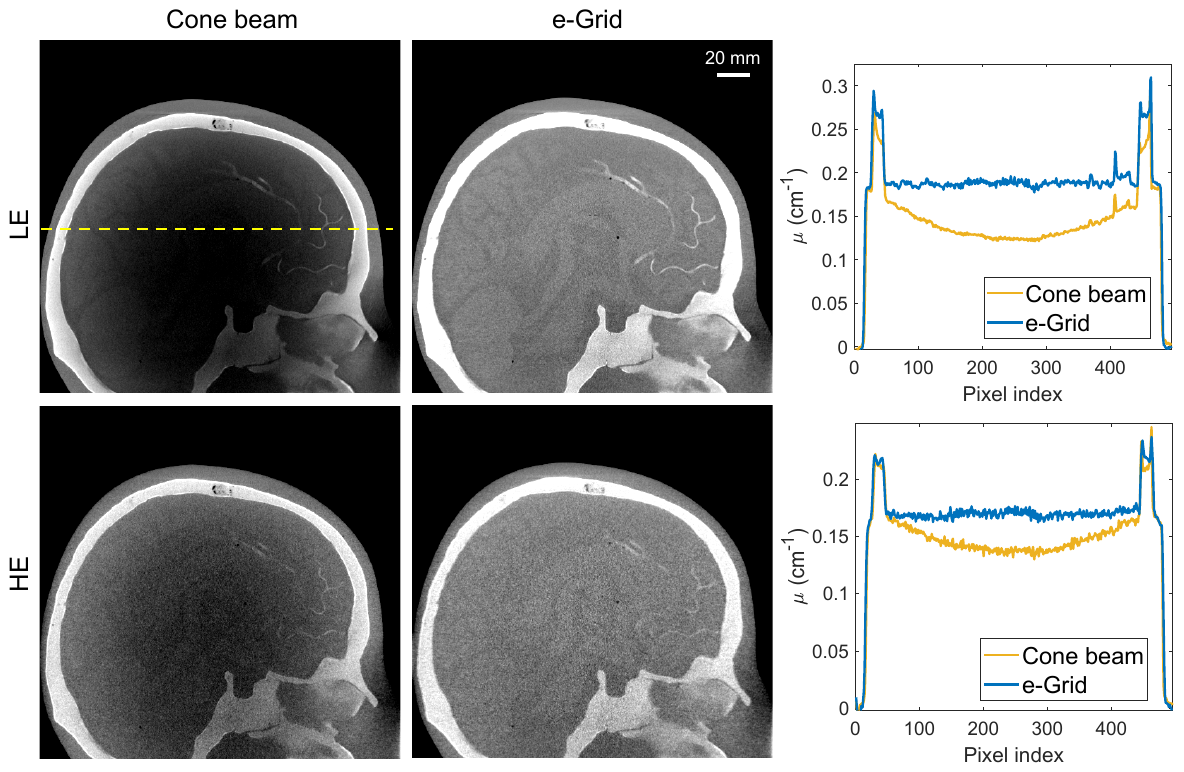}
   \caption{Imaging results of head phantom on the sagittal view plane. The horizontal line profiles were presented on the right-hand side. The display window was $\left[0.12, 0.25\right]$ $\rm{cm}^{-1}$ for low-energy CT images, and $\left[0.12, 0.22\right]$ $\rm{cm}^{-1}$ for high-energy CT images. The scale bar denoted 20 mm.} 
   \label{fig:sagittal}
    \end{center}
\end{figure}
The experimental results of the head phantom were shown in Fig.~\ref{fig:sc_results_KYOTO} and Fig.~\ref{fig:sagittal}. Clearly, distinguishing the brain tissue was quite challenging before correcting the scatter shading artifacts. Again, results demonstrated that the developed e-Grid method can effectively suppress the shading artifacts for both low-energy and high-energy CT images, see Fig.~\ref{fig:sagittal}. As a result, the image quality and readability were greatly enhanced after processed by the e-Grid method.

\begin{figure}[htbp]
	\begin{center}
		\includegraphics[width=1\linewidth]{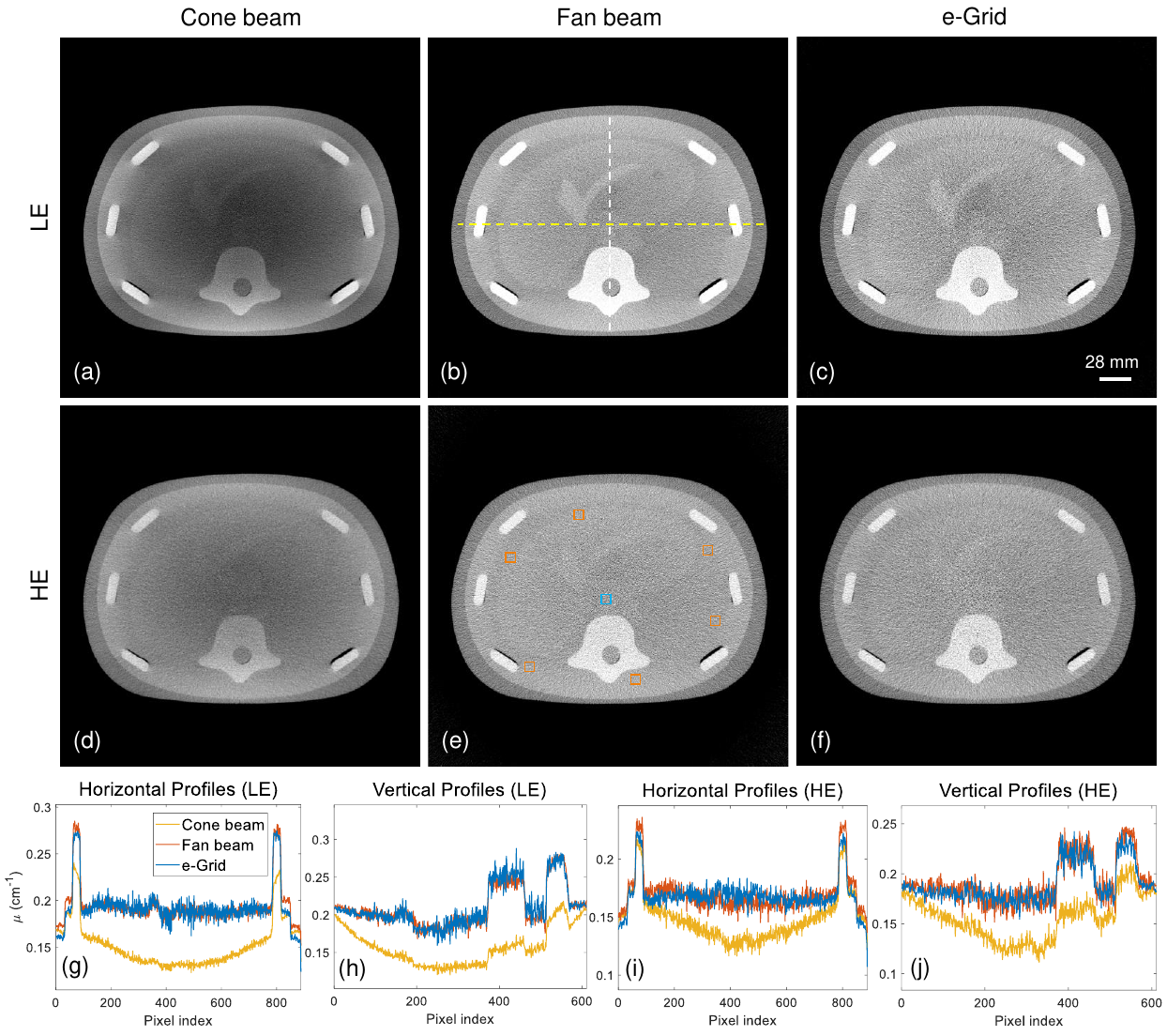}
		\caption{Imaging results of the abdominal phantom. The profiles along the horizontal and vertical directions were compared and plotted at the bottom. The display window was $\left[0.10, 0.25\right]$ $\rm{cm}^{-1}$ for low-energy CT images, and $\left[0.07, 0.24\right]$ $\rm{cm}^{-1}$ for high-energy CT images. The scale bar denoted 28 mm.}
		\label{fig:sc_results_abdominal}
	\end{center}
\end{figure}

The experiment results of the abdominal phantom were shown in Fig.~\ref{fig:sc_results_abdominal}. Different from the water phantom and head phantom, the abdominal phantom had a larger diameter of 28 cm. Similarly, distinguishing tissues in the central region of the abdominal phantom before scatter correction was challenging. Reducing the beam width can help alleviate such shading artifacts. The e-Grid method was able to effectively mitigate the scatter artifacts in the low-energy and high-energy CT images of the abdominal phantom, see Fig.~\ref{fig:sc_results_abdominal}(c) and Fig.~\ref{fig:sc_results_abdominal}(f). These above results indicated that the proposed e-Grid method can be applied to correct the scatter shading artifacts of different object sizes, provided that proper calibration parameters were employed.

\begin{figure}[htbp]
   \begin{center}
   \includegraphics[width=0.9\linewidth]{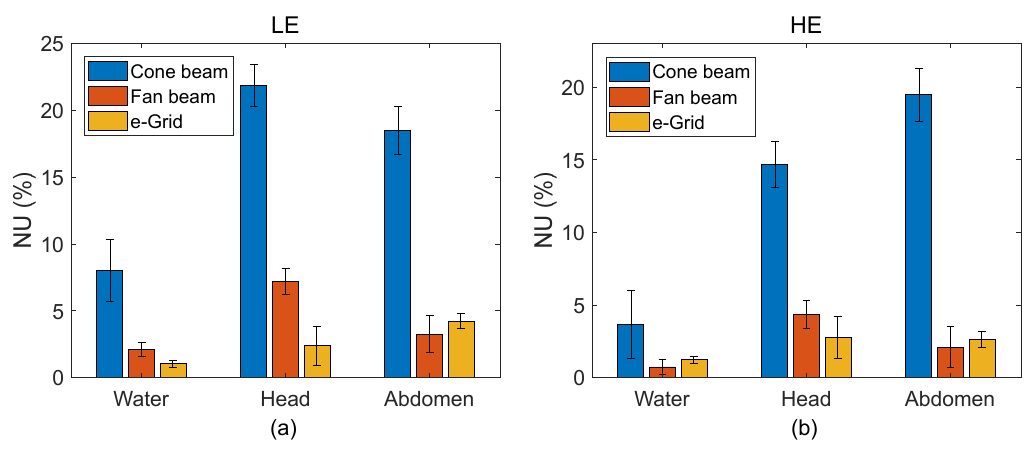}
   \caption{The measured non-uniformity (NU) indices on the (a) low-energy CT images and (b) high-energy CT images for the water phantom, head phantom and abdominal phantom, respectively.}
   \label{fig:numu}
    \end{center}
\end{figure}

Besides, image uniformity were measured, see the statistical results in Fig.~\ref{fig:numu}. The selected ROIs for the water phantom, the head phantom and the abdominal phantom were shown in Fig.~\ref{fig:sc_results_insert}(e), Fig.~\ref{fig:sc_results_KYOTO}(e) and Fig.~\ref{fig:sc_results_abdominal}(e), respectively. The blue-box area represented the central ROI, while the six orange-box areas represented the peripheral ROIs. Due to the variations of the phantom size and the ROI positions, slight differences in NU values were observed among the three phantoms. Compared to the results obtained from the cone beam imaging setup, the NU value measured on the e-Grid processed low-energy CT images was reduced by over 87\%, 90\% and 77\% for the water phantom, the head phantom, and the abdominal phantom, respectively. In addition, the NU value measured on the e-Grid processed high-energy CT images was reduced by over 66\%, 81\% and 86\% for the water phantom, the head phantom, and the abdominal phantom, respectively.

Finally, the measured SNR results were shown in Table~\ref{tab:snr}. The same ROIs were used for SNR measurements. Compared to the SNR before scatter correction, the SNR measured on the e-Grid processed low-energy CT images decreased by approximately 8\%, 21\% and 27\% for the water phantom, the head phantom, and the abdominal phantom, respectively. The SNR values measured on the e-Grid processed high-energy CT images decreased by approximately 4\%, 14\% and 18\% for the water phantom, the head phantom, and the abdominal phantom, respectively. One should be aware that the SNR reduction was mainly due to the removal of the scattered X-ray photons\cite{wang2010scatter,niu2011scatter}.

\begin{table}[!htb]
\begin{center}
\caption{The measured SNR values for low-energy and high-energy CT images.
\vspace*{2ex}
}
\renewcommand\arraystretch{1.3}
\resizebox{0.8\columnwidth}{!}{
\begin{tabular} {l@{\hspace{0.5cm}}c@{\hspace{0.5cm}}c@{\hspace{0.5cm}}c@{\hspace{0.5cm}}c}
\toprule[1pt]
\hline
&Cone beam(LE)&e-Grid(LE)&Cone beam(HE)&e-Grid(HE)\\
\cline{1-5}
Water phantom& 54.23$\pm$5.07 & 49.65$\pm$4.28 & 33.72$\pm$1.90 & 32.53$\pm$1.75 \\
Head phantom& 48.72$\pm$4.47 & 38.77$\pm$5.16 & 34.23$\pm$5.27 & 29.27$\pm$3.55 \\
Abdominal phantom& 24.69$\pm$2.57 & 17.87$\pm$2.63 & 19.54$\pm$2.62 & 16.08$\pm$3.04 \\

\hline
\bottomrule[1pt]
\end{tabular}}
\label{tab:snr}
\end{center}
\end{table}

\begin{figure}[htbp]
   \begin{center}
   \includegraphics[width=1\linewidth]{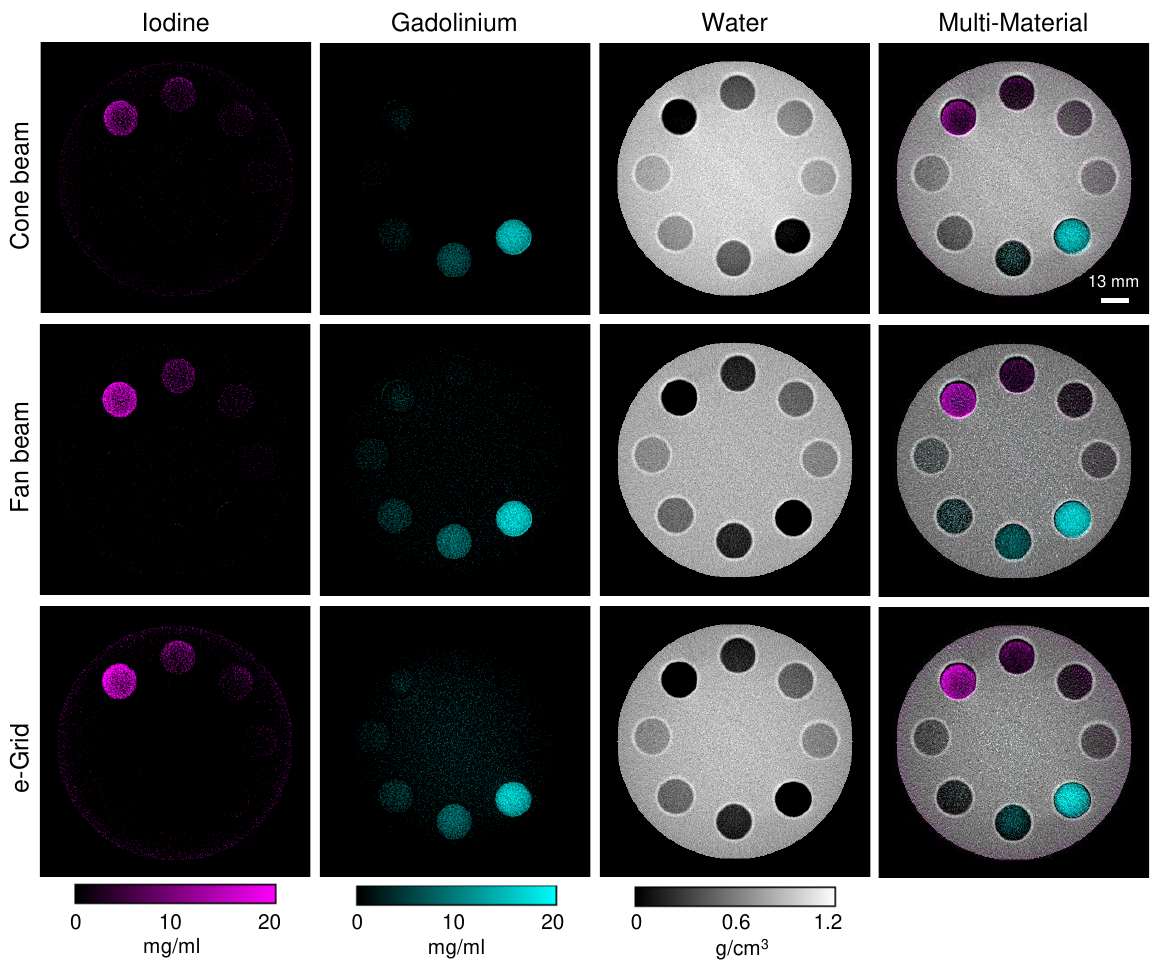}
   \caption{The multi-material decomposition results of the water cylindrical phantom. From top to bottom, results correspond to the cone beam, fan beam and e-Grid corrected CT images. From left to right, images correspond to the iodine, gadolinium, water bases and color overlaid images. The display window was $\left[0, 20\right]$ $\rm{mg/ml}$ for iodine and gadolinium maps, and $\left[0, 1.2\right]$ $\rm{g/cm^3}$ for water maps. The scale bar denoted 13 mm.}
   \label{fig:decomposition}
    \end{center}
\end{figure}

\subsection{Decomposition results}
Material specific maps were generated from the low-energy and high-energy CT images using the multi-material decomposition algorithm (MMD) \cite{mendoncca2013flexible}, see the results in Fig.~\ref{fig:decomposition}. From top to bottom, results correspond to the uncorrected cone beam setup, the fan beam setup, and the e-Grid method. The decomposed iodine basis was marked in magenta, and the decomposed gadolinium basis was marked in cyan. Overall, the accuracy of the material bases before correction was low, see the measured material densities in Table~\ref{tab:dec}. The number in parentheses indicated the ground truth. The quantitative decomposition results of the iodine basis with the e-Grid method closely match the ground truth, demonstrating its enhanced performance in quantitative DE-CBCT imaging.

\begin{table}[htbp]
\begin{center}
\caption{Quantitative decomposition results of the water phantom.
\vspace*{-2ex}
}
\renewcommand\arraystretch{1}
\resizebox{1\columnwidth}{!}{
\begin{tabular} {lcccccccccc}
\toprule[1pt]
\hline
\multicolumn{3}{c}{Iodine (mg/ml)} &&\multicolumn{3}{c}{Gadolinium (mg/ml)}&&\multicolumn{3}{c}{Water ($\rm{g/cm^3}$)}\\
\cline{1-3} \cline{5-7} \cline{9-11} 
Cone beam&Fan beam&e-Grid&&Cone beam&Fan beam&e-Grid&&Cone beam&Fan beam&e-Grid\\
\cline{1-11}
12.823(20)&18.651(20)&19.568(20)&&19.521(20)&21.085(20)&19.866(20)&&1.159(1)&1.033(1)&1.062(1) \\
7.471(10)&9.406(10)&10.479(10)&&8.640(10)&10.762(10)&10.342(10)&&-&-&- \\
4.382(5)&4.810(5)&5.866(5)&&3.259(5)&5.769(5)&5.249(5)&&-&-&-  \\
2.648(2.5)&2.093(2.5)&3.032(2.5)&&0.636(2.5)&3.308(2.5)&2.883(2.5)&&-&-&-  \\
\hline
\bottomrule[1pt]
\end{tabular}}
\label{tab:dec}
\end{center}
\end{table}

\section{Discussions}\label{sec: discussions}
This study presented a novel scatter correction method, named as e-Grid, for DL-FPD based CBCT imaging. With DL-FPD, two sets of primary and scattered X-ray signals can be measured independently at two distinct energy levels. Under a linear approximation, the low-energy and high-energy scattered signals can be estimated analytically. Consequently, scatter corrections can be performed easily for the dual-energy data acquired from the DL-FPD without the need of additional hardware such as anti-scatter grids. To validate the newly proposed e-Grid method, Monte Carlo simulation experiments and physical phantom experiments were conducted. Results demonstrated the high performance of the e-Grid scatter correction method in reducing the shading artifacts on the CBCT images. Quantitative analyses showed that the e-Grid method was able to reduce the image non-uniformity by over 90\% and 60\% for the low-energy and high-energy CT images, respectively.

The assumptions made in Eqs.~(\ref{eq:3})-(\ref{eq:4}) were fundamental to the e-Grid method. An implicit requirement underlying Eqs.~(\ref{eq:3})-(\ref{eq:4}) was that the acquired low-energy and high-energy signal were spectrally distinct. With DL-FPD, such requirement can be easily satisfied, especially when an additional 1.0 mm thick copper filter was placed between the top and bottom detector layers. Additionally, this newly proposed e-Grid scatter correction method may also be applied in other dual-energy CBCT imaging systems that utilize dual X-ray sources and detectors. For that specific dual-energy CBCT imaging setup, the acquired projections from the two individual source-detector systems should be accurately registered to minimize potential geometric inconsistencies. Finally, this proposed e-Grid scatter correction method may also be applicable to triple-layer FPDs.

The current study may have some limitations. Firstly, a group of pre-calibrated parameters $\alpha^p_{1}$, $\alpha^p_{0}$, $\alpha^s_{1}$ and $\alpha^s_{0}$ were valid only for objects having dimensions similar to those of the calibration phantom. For example, the parameters calibrated from a 16 cm PMMA phantom were valid for head imaging, whereas the calibrated parameters from the 32 cm PMMA phantom were valid for body imaging. To image objects of other sizes, additional calibration experiments with phantoms of specific sizes should be conducted. Secondly, it had been observed that the removal of scatter signals leads to a slight increase in CT image noise. This occurred because subtracting the scattered signals reduced the total number of X-ray photons. However, the decrease in SNR was subtle compared to the substantial improvements in image quality and the accuracy of quantitative material decomposition. For example, the SNR of the water phantom decreased by 8\% for low-energy CT images and by 4\% for high-energy CT images. Whereas, the NU value decreased by 87\% for low-energy CT images and 66\% for high-energy CT images. Thirdly, it would be interesting to investigate the scatter correction performance of the e-Grid approach with MeV X-ray beams in future studies with respect to applications of megavoltage image-guided radiation therapy\cite{spies2001correction, maltz2008algorithm, boylan2012megavoltage}. Fourth, the performance of e-Grid for objects with larger diameters, e.g., $\ge 40$ cm, need to be investigated in the future on a DL-FPD based CBCT system. Our experience suggests that it is generally safe to use the same set of correction coefficients for objects with size deviations of less than 5 cm. Therefore, certain calibration experiments are necessary to acquire the corresponding correction parameters to generate the most optimal imaging performance for different sized objects. For practical applications, we think a dedicated lookup table might be beneficial.

\section{Conclusion}\label{sec: conclusion}
In conclusion, a novel scatter correction method, named as e-Grid, is proposed for DL-FPD based dual-energy CBCT imaging. It can quickly estimate the scattered signals from the acquired low-energy and high-energy projections. Experiments demonstrate that the e-Grid method can effectively reduce the shading artifacts, thereby significantly improving the quality of CBCT images and the accuracy of material decomposition. In the future, scatter artifacts might be easily corrected for the DL-FPD based dual-energy CBCT imaging systems.


\bibliography{./Bibliography_Paper}
\bibliographystyle{./medphy}
\end{document}